# A Privacy-Preserving Ecosystem for Developing Machine Learning Algorithms Using Patient Data: Insights from the TUM.ai Makeathon

Simon Süwer[1,*], Mai Khanh Mai[2,*], Christoph Klein[2,3], Nicola Götzenberger[2], Denis Dalić[5], Andreas Maier[1,†], Jan Baumbach[1,4,†]

[1] Institute for Computational Systems Biology, University of Hamburg, Hamburg, Germany

[2] Dr. von Hauner Children's Hospital, LMU Klinikum, Ludwig-Maximilians-Universität München, Munich, Germany

[3] German Center for Child and Adolescent Health (DZKJ), Munich Site, Munich, Germany

[4] Computational Biomedicine Lab, Department of Mathematics and Computer Science, University of Southern Denmark, Odense, Denmark

[5] MI4People – Non-Profit for AI for Social Good, Munich, Germany

[*] These authors are first authors of the paper

[†] These authors supervised the project

## Abstract

The integration of clinical data offers significant potential for the development of personalized medicine. However, its use is severely restricted by the General Data Protection Regulation (GDPR), especially for small cohorts with rare diseases. High-quality, structured data is essential for the development of predictive medical AI.

In this case study, we propose a novel, multi-stage approach to secure AI training: (1) The model is designed on a simulated clinical knowledge graph (cKG). This graph is used exclusively to represent the structural characteristics of the real cKG without revealing any sensitive content. (2) The model is then integrated into the FeatureCloud (FC) federated learning framework, where it is prepared in a single-client configuration within a protected execution environment. (3) Training then takes place within the hospital environment on the real cKG, either under the direct supervision of hospital staff or via a fully automated pipeline controlled by the hospital. (4) Finally, verified evaluation scripts are executed, which only return aggregated performance metrics. This enables immediate performance feedback without sensitive patient data or individual predictions, leaving the clinic. A fundamental element of this approach involves the incorporation of a cKG, which serves to organize multi-omics and patient data within the context of real-world hospital environments.

This approach was successfully validated during the TUM.ai Makeathon 2024 (TUMaiM24) challenge set by the Dr. von Hauner Children's Hospital (HCH-LMU): 50 students developed models for patient classification and diagnosis without access to real data.
Deploying secure algorithms via federated frameworks, such as the FC framework, could be a practical way of achieving privacy-preserving AI in healthcare.

# Introduction

The substantial increase in data within the clinical environment, consequent to declining diagnostics costs, presents a distinctive opportunity to exploit the potential of big data and advanced analytics to help current and future patients by gaining novel insights into poorly understood disorders [1,2]. This growth in data availability has driven innovation in medical research, particularly in pediatric health [3]. Nevertheless, stringent data protection regulations, such as the GDPR, impose considerable restrictions on access to high-quality patient data, thereby significantly challenging the research process. While these measures are crucial for protecting sensitive patient data, they also slow down the further development of AI-supported innovations [4–6]. Personalized medicine is a prime example of this, as a field that depends on utilizing diverse and comprehensive data sets to develop reliable models. In the absence of public accessibility to these data, the training of these models can be undertaken within the confines of hospital infrastructure or data repositories [7]. This challenge is particularly evident in the study of rare diseases. In the EU, a disease is classified as rare if it affects fewer than five individuals per 10,000 inhabitants [8,9]. Even with an increase in clinical data, the patient population of rare diseases is limited and becomes even smaller when focusing exclusively on pediatric cases [10]. Therefore, the rarity of individuals with the same disorder highlights the need for data sharing to establish a larger research pool [11]. This creates a fundamental dilemma: While data on children is necessary to improve research and ultimately children's health [12], there is a legitimate interest in protecting their data privacy [13].
Statistics reveal that parents exhibit greater reluctance to share data for their underage children than for themselves [14,15]. However, they are more likely to consent to the use of their children's data if the security of the data and its integrity can be guaranteed [14]. One easy way of getting parental consent is to store data exclusively and securely in one hospital system and ensure that sensitive information does not leave the primary facility [16].
Hospitals such as the Dr. von Hauner Children's Hospital routinely analyze complex multi-omics datasets on rare pediatric diseases for diagnostic purposes. However, due to limited resources, such institutions often face challenges in conducting further in-depth research on these diseases [17,18]. While data collection can be carried out by experts in specialized clinics, the limitations of the GDPR prevent the same flexible approach from being applied to data access and analysis. This is because data cannot simply be collected or shared centrally, which hinders the development of innovative algorithms and limits the possibilities of utilizing state-of-the-art techniques such as machine learning (ML) [19] (Figure 1A).
A promising approach to overcome data protection and privacy issues in healthcare is Federated Learning (FL) [20] (Figure 1C). Unlike traditional ML models, which require data to be aggregated in a single location, FL enables institutions to train models via a decentralized process, whilst preserving data privacy [4,20]. This approach ensures that sensitive patient data remains within the hospital walls, enabling organizations to train models collaboratively without disclosing raw data and thus complying with data protection regulations [4,20]. In FL, local datasets from participating institutions are used to create local AI models. After each training round, the model parameters are aggregated to create a global model that is trained on all local data sets without the need to share any sensitive patient data [21,22]. A substantial corpus of research has demonstrated that FL has the capacity to integrate local training with global model optimization, thus presenting a promising approach for data protection and enhancing healthcare efficiency [20,23–25]. For example, FL has been used to successfully detect cardiovascular disease by combining ECG and echocardiography data from multiple institutions [26]. In addition, the ICU4Covid project developed an FL-based decision support system that enables early identification of high-risk hypertensive patients while maintaining data protection [7]. Furthermore, the benefits of FL in predicting the risk of sepsis and kidney injury have been demonstrated in a practical setting [27]. In addition, by offering computing capacity on hospital data sets without disclosing the data itself, hospitals can generate additional revenue that can be reinvested in advanced diagnostic methods such as genomics and proteomics. For pharmaceutical companies and software developers, FL provides a way to identify new clinical or biological markers for drug development, drug repurposing, or stratification algorithms [28]. However, a notable limitation of learning algorithms is that the data structure must be known prior to their development [20].

While these examples highlight the strong potential of FL in multi-institutional healthcare settings, current implementations largely rely on collaborations across multiple data providers. This raises a critical and largely unexplored question:

Can FL be utilized in a single-client configuration to enable ML developers to build and test their own FL algorithms while remotely accessing patient data?

This study investigates the use of an FL framework for secure single-client model training as a privacy-preserving paradigm for developing algorithms for ML in clinical research. As a proof of concept, the Federated Learning platform FeatureCloud (FC) was tested by the Dr. von Hauner Children's Hospital during the TUM.ai Makeathon 2024 (TUMaiM24) [21] (Figure 1B). Additionally, it was evaluated how well participants can develop model architectures for patient identification and disease type classification based on patient data stored in a clinical knowledge graph (cKG) without ever accessing the real data or training the models themselves. Instead, participants were given an artificial cKG filled with synthetic data mimicking the structure of the real patient data. Leveraging the privacy-preserving FC platform ensured that all training steps and evaluation computations took place locally on the hospital infrastructure, with only model updates exchanged with the external developers. This approach protects data privacy by keeping sensitive information secure, while allowing participants to design models aligned with the real clinical data environment. cKGs enable the integration of different data sources, for example, in the field of multi-omics analyses, and are already being tested for disease classification and therapy recommendations [29].

Recent studies have demonstrated an increasing focus on approaches that integrate federated learning with cKGs [30–33].

Participants had the opportunity to use synthetic data to create their model architectures without having to disclose real, confidential patient data. In this case, the simulated cKG was not used to train the models, but exclusively to replicate the structural properties of the real knowledge graphs. As a result, it was possible to design models that are compatible with real data without ever needing access to it. In the long term, the use of more advanced synthesis methods opens up the potential to generate more realistic scenarios and thus further improve the validation and robustness of models [34–36]

Verifying that FL can be utilized in a single-client configuration is particularly pertinent in scenarios where data protection is essential and only the structure of data is known.
While the primary goal of FL is to enable privacy-preserving collaboration across multiple parties [37], deploying FL infrastructure in a single-client configuration offers a valuable development pathway. It allows ML developers to build and test their own FL algorithms securely, with remote access to sensitive data remaining protected. Importantly, such setups are inherently scalable, providing a foundation that can later be extended to full multi-client, privacy-preserving model training.
To the best of the authors' knowledge, there is no other research on this specific use case.

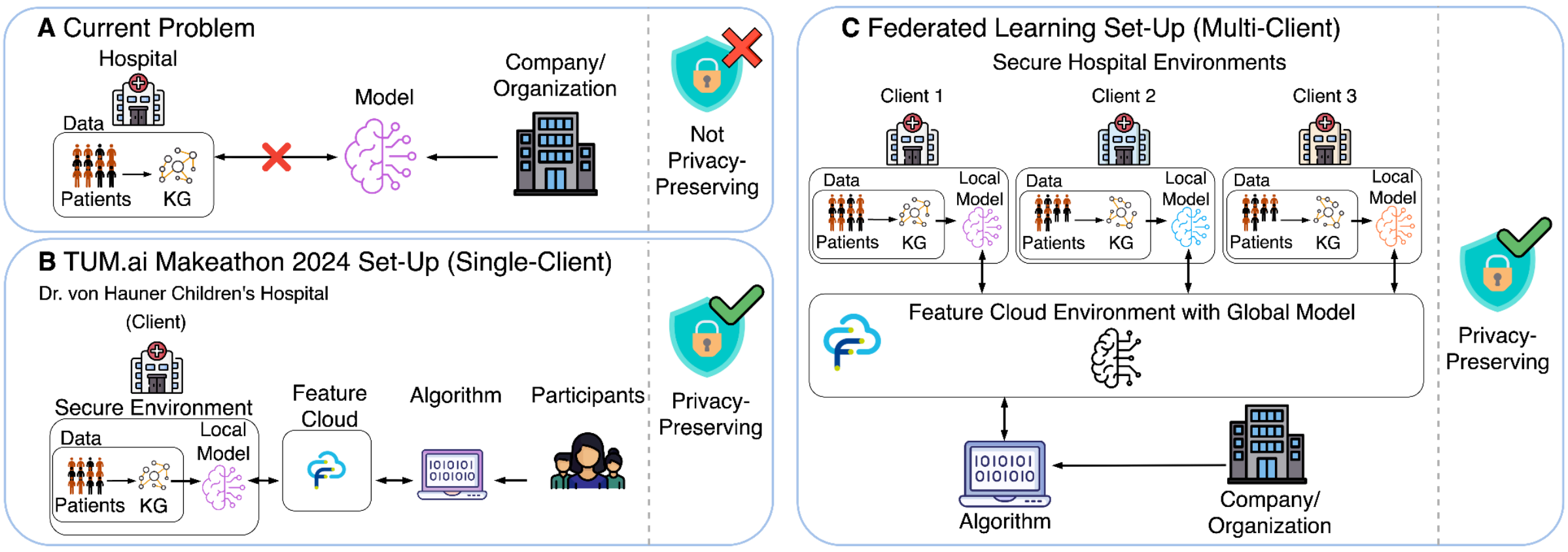

**Figure 1 - Overview of the limits of privacy-preserving AI model training (A), Test FL Set-Up in a single-client scenario (B), and multi-client FL Set-Up (C)**: The overall problem is having relevant patient data but not the means to securely allow research on this data (A). To compensate for this, the goal would be to leverage FeatureCloud as an FL approach (C). This enables other research facilities (e.g., companies/organizations) to submit an ML algorithm, which can be trained via FeatureCloud by training local models based on their local knowledge graph (KG) in each participating hospital (clients) and only exchanging parameters with the global model, preserving patient data privacy. A simplified version using a single client scenario was tested in the TUM.ai Makeathon 2024, where

participants provided an ML algorithm to train with the data in the secure environment of the Dr. von Hauner Children's Hospital (B).

# Methods

This case study investigated the use of FL in a single-client setting. In this scenario, the Dr. von Hauner Children's Hospital acts as the data-holding client, and external developers interact with the data only through the federated system without any direct access. This challenge was part of the TUM.ai Makeathon 2024, an annual 36-hour programming hackathon offering various challenges for its participants to explore the practical use of AI methods in real-world settings. The clinical challenge was provided by Dr. von Hauner Children's Hospital in collaboration with other partners, with approximately 30 participants working on it to test the study's concept (Figure 1B).

## Knowledge Graphs

To ensure data protection, the participating developers were given synthetic data in the form of a simulated knowledge graph $KG_s$ mimicking the data structure of the original knowledge graph $KG_o$. $KG_o$ is the graph maintained at the Dr. von Hauner Children's Hospital and is based on an extended and adapted version of the Clinical Knowledge Graph $KG_c$, an open-source platform comprising close to 20 million nodes and 220 million relationships representing relevant experimental data, public databases, and literature [29]. $KG_o$ contains genetic patient data, where each patient is represented as a subject node. Each subject has one or multiple biological samples drawn from the patient at different time points (e.g., blood, urine, or plasma samples). These samples are linked with respective genes and proteins. Additionally, if a clinical diagnosis is available, the sample is also linked to the corresponding disease(s). Both $KG_s$ and $KG_o$ have the same set of node and edge types (Figure 2).

The data for the $KG_s$ was taken from the MIMIC-III Demo Dataset [38], an open-source subset containing information on 100 patients, including their diagnoses and lab results [38,39]. Consequently, for the $KG_s$ 100 Subject nodes were created, each representing a patient. In the simulated graph, every Subject node is linked to a single Biological_Sample (BS) node via the BELONGS_TO_SUBJECT relationship, simulating the collection of one biological sample per patient.
Each diagnosis from the MIMIC-III Dataset was represented as a Disease node, extending the already existing Disease nodes from the $KG_o$ and resulting in a total of 10792 Disease nodes in the $KG_s$. Similarly to the $KG_o$, there is one Disease node labeled as "control" for the cohort of healthy patients. Among the 100 BS nodes, 99 were assigned to their corresponding disease diagnosis, while the remaining one was linked to the control, each via a HAS_DISEASE relationship. If an ICD-10 and/or ICD-9 code was available for a diagnosis, it was incorporated into the “synonyms” property of the respective Disease node.
Furthermore, lab results were translated into Human Phenotype Ontology (HPO) terms using the LOINC2HPO tool [40]. Each HPO term was then stored as a Phenotype node in the $KG_s$.
To model relationships between BS and Gene nodes, each BS was randomly assigned to genes via a HAS_DAMAGE relationship. Each connection was assigned a random score between 1 and 20, simulating variant deleteriousness as measured by the CADD score [41]. Similarly, BS nodes were randomly linked to Protein nodes through a HAS_PROTEIN relationship, with a random score between 1 and 20 mimicking log10 quantification scores obtained via mass spectrometry. On average, each biological sample in the $KG_s$ is associated with approximately five randomly assigned genes and 50 randomly linked proteins.

Storing genetic patient data within a knowledge graph offers a powerful advantage for developers by providing a structured, semantically rich, and easily navigable representation of complex biological relationships. Unlike traditional databases like DISEASES [42,43] or DrugBank 5.0 [42,43], knowledge graphs capture the intricate links between patients, genes, phenotypes, and diseases, enabling developers to query and explore meaningful biological pathways without needing deep domain knowledge. This interconnected format allows for more intuitive feature extraction, supports the development of explainable machine learning models, and simplifies the integration of additional biomedical data sources [44]. In this context, it empowers developers to build advanced algorithms while working within a privacy-preserving environment. Stored in a Neo4j database, the $KG_s$ graph structure was locally accessible to participants (Figure 2), allowing them to explore its schema and relationships. Based on this structure, they were tasked with developing model applications capable of extracting the relevant information needed for training their machine learning models.

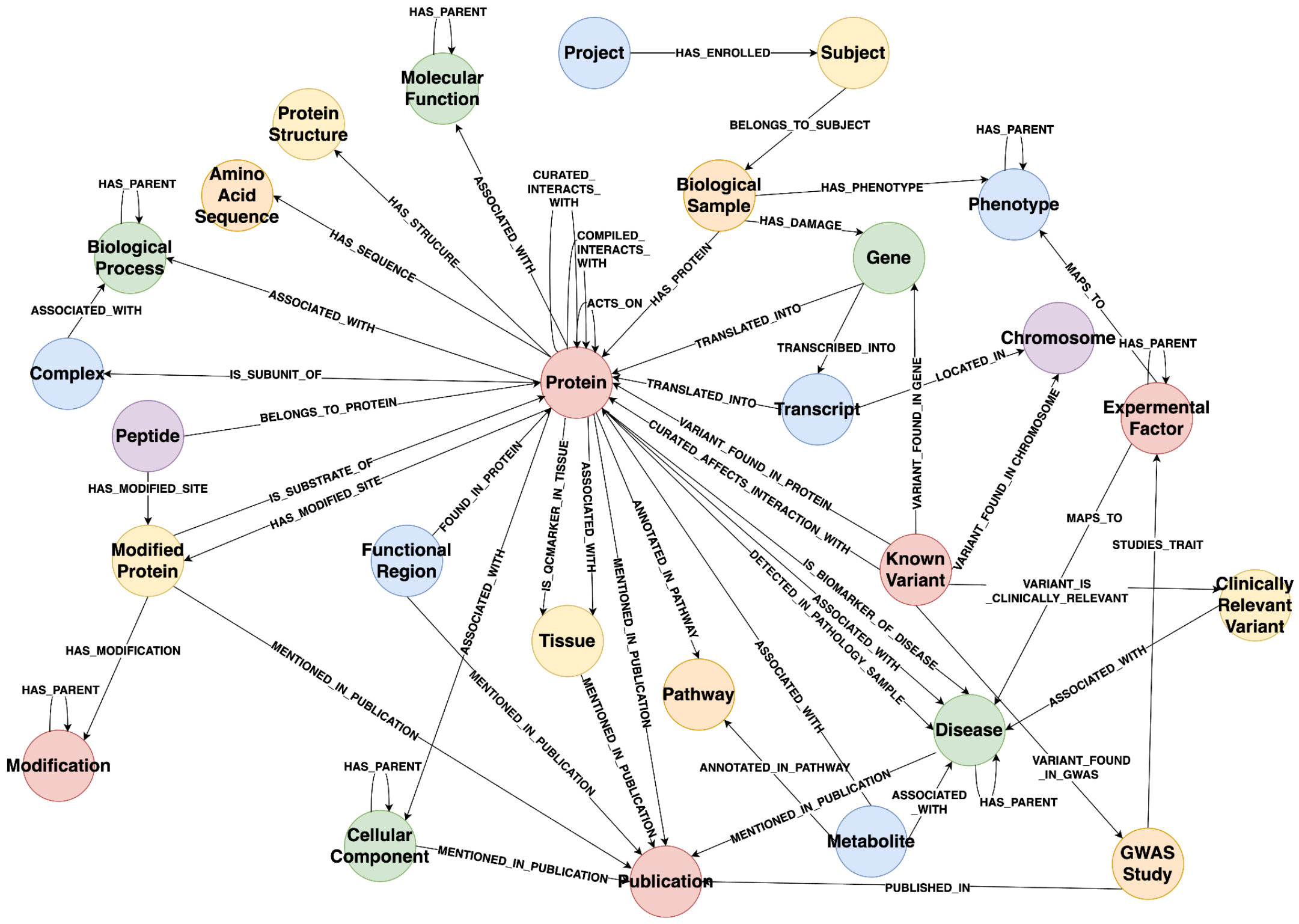

**Figure 2 - Overview of the Knowledge Graph Structure**: It represents the nodes and their relations in the $KG_o$ and $KG_s$.

The utilization of $KG_s$ during the TUMaiM24 has enabled the development of ML models that can be employed in the Dr. von Hauner Children's Hospital without compromising the integrity of data protection regulations. The underlying structure of the real data remains intact, ensuring that models trained on synthetic data can also be effectively deployed on actual patient data. Notably, while synthetic data replicates the structure of real patient data, it does not contain any personal information, thereby enhancing its security and compliance with data protection regulations. Building upon this foundation, the TUMaiM24 participants endeavored to engineer models capable of distinguishing between healthy and sick patients, in addition to predicting the first letter of ICD-10 codes.

## Execution Environment

In the context of the TUMaiM24 challenge, participants were tasked with the development of machine learning models on a virtual machine operating Ubuntu 22.04.5 LTS. The virtual machine was configured with 70 virtual CPUs and 100 GB of RAM, thereby ensuring sufficient computing power for the execution of complex training processes. To preserve the confidentiality of patient data, participants were not granted direct access to $KG_o$. Instead, they were provided with $KG_s$ for the development of their models and only received aggregated statistics of their models trained with the real data. The original data is stored in a Neo4j database[1] that could be containerized using a Docker[2] image based on Ubuntu 22.04 and was made available in a provided GitHub repository. This allowed participants to experiment with synthetic demo data locally without ever seeing sensitive patient data [29]. The participating teams used the FC infrastructure to submit their respective models, with a Docker container providing encapsulation of the applications. Following this, authorized personnel within the hospital environment initiated the process using an FC workflow to access the actual patient data. After the completion of each training cycle, the participants received statistical key figures, including accuracy, precision, and recall, thereby enabling the optimization of their respective models in subsequent steps. Notably, these calculations were not performed by the participants' applications; instead, they relied on an independent script, which was executed manually by the challenge staff (Figure 3).

The hospital stored all sensitive information and employed firewalls to protect it. The transfer of model parameters rather than raw data led to a significant reduction in the risk of data leaks. Furthermore, the utilized Docker containers were meticulously examined prior to deployment to ensure adherence to all security and data protection requirements. These measures established a connection between advanced research in machine learning and stringent requirements for patient data protection. Concurrently, a reproducible and transparent process was implemented for all participants.

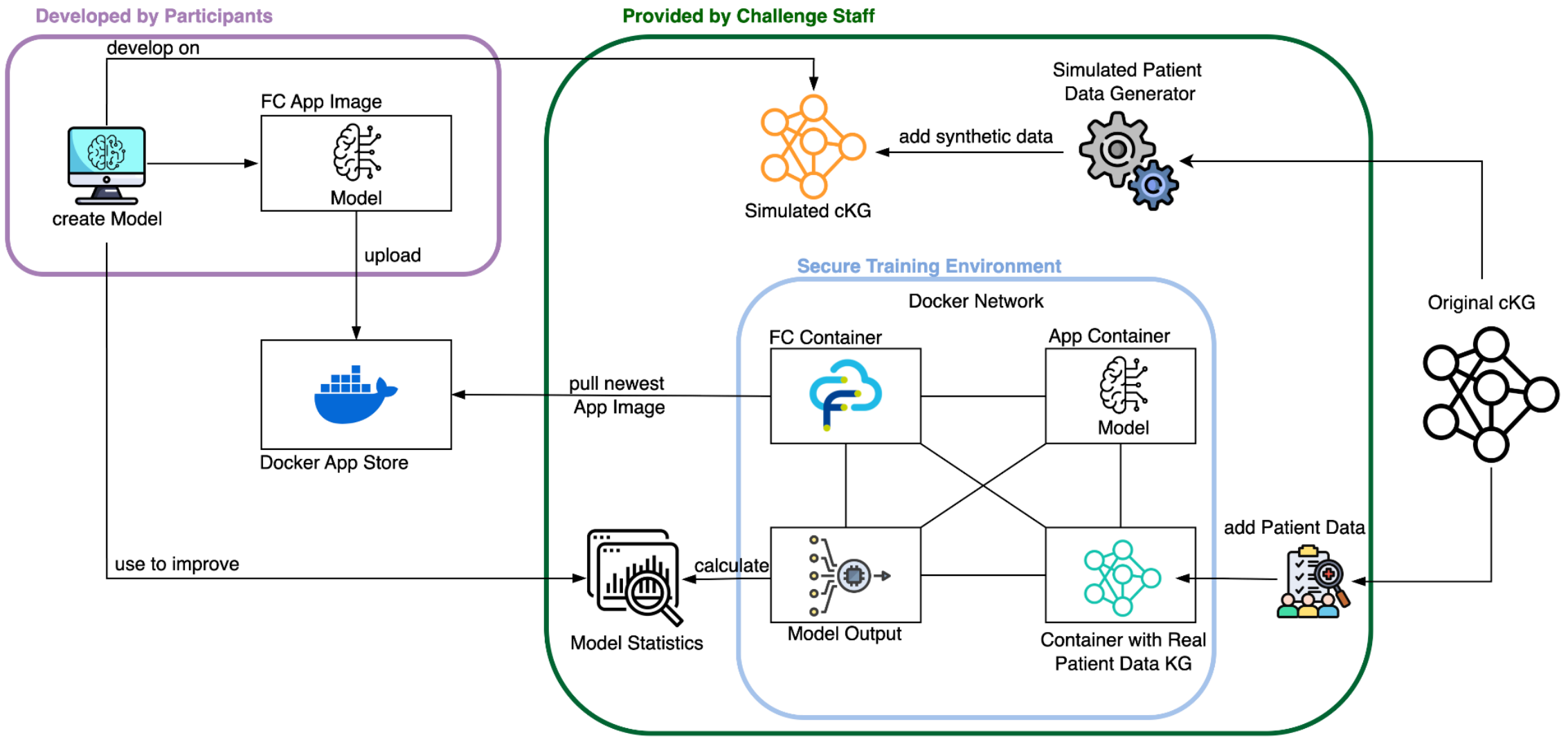

**Figure 3 - Makeathon Data Workflow:** Participants developed a model using data from the simulated cKG and integrated it into an FC app image. This image is then uploaded to the FC App Store, allowing the FC Controller to pull the latest model version and create an App Container containing the model. Within a Docker Network, the model container interacts with the real KG, which contains actual patient data, to train the model via the FC Controller. After training, the model is evaluated by generating statistics to assess model performance. Participants can use these insights to refine their model and repeat the process until

[1] https://neo4j.com/
[2] https://www.docker.com/

submitting the final version. Tasks assigned to participants are highlighted in purple, while resources provided by developers and challenge staff are shown in green/blue. The simulated cKG is based on the Clinical Knowledge Graph [29] with synthetic data ($KG_s$), while the original cKG shares the same structure but with real patient genetics gathered and stored in the Dr. Hauner Children's Hospital ($KG_o$).

## TUM.ai Makeathon 2024 Clinical Challenge

The TUMaiM24 provided a unique opportunity for students and young professionals to engage in developing ML models for medical research. Participants were given two tasks[3]. For each task, the participants should create a model that utilizes $KG_o$ as a data source.

- **Task A:** The first task was to classify patients as diseased or healthy. For this, all patient nodes *Biological_samples* that are connected to a *Disease* node via a *HAS_DISEASE* relationship are used. The control group corresponds to the *Disease* node labeled *control*, while a subset of unconnected *Biological_samples* is reserved as a validation set.
- **Task B**: The second task was to develop a model that predicts the first letter of the International Classification of Diseases (ICD) code to identify the general disease category of the patients. For training, all *Biological_samples* are used that are connected to a *Disease* node via a *HAS_DISEASE* relationship, excluding *Disease* nodes labeled *control,* so that only diseased patients are included in the training group. The ICD-10 is annotated under the *synonyms* property in the *Disease* node.

The result of both tasks required output in tabular form, with the first column containing the *subject_id* from the *biological_samples*. The second column contains either the predicted disease status, where 0 stands for control samples and 1 for diseased individuals (Task A), or the first letter of the ICD code used to classify the general disease category (Task B). The output must be saved as a CSV file. Based on this CSV, a separate evaluation was carried out in the controlled environment, and the scores achieved were reported back to the participants as evaluation metrics. While the classification in Task A is evaluated using the $F_1$ score, the evaluation of the model in Task B is based on the recall value. The F1 score (Task A) is regarded as the harmonic mean, which serves to optimize the trade-off between the two. This is particularly salient for binary classifications such as "sick" vs. "healthy." Conversely, recall (Task B) is imperative for disease diagnosis because it maximizes the identification of diseased cases [45]. The formal definition of these metrics can be found in the supplements. The competitors who achieved the highest total score were declared the winners of the challenge. This total score was determined by the sum of the scores obtained for *Task A* and *Task B*, with a maximum possible score of 2. In total, around 30 participants contributed to the challenge in groups of a maximum of four members.

[3] https://tum-ai.notion.site/AI-in-Medicine-Project-AMIGO-d882e781fbfc4056b474fee54cdb0b2f

# Experimental Results

## Makeathon Test-Run of the Secure Training Pipeline

During the short time span of 36 hours in the TUMaiM24, four models were successfully trained and submitted using the FC framework for Task A. For Task B, three models were submitted, of which two models achieved a score higher than zero (see Table 1). Given the limited time, the algorithms developed were simple, mainly using random forest techniques, and did not achieve high scores. For Task A, the highest $F_1$ score was 0.67, and for Task B, the highest recall was 0.39 (see Table 1).

Table 1: The table shows the top 4 most successful teams. The teams are listed from highest to lowest total score, with the winning team in the first row. Bold values mark the highest score for the category (column). A dash (-) indicates that the group has not completed the task. Their scores per task

| | Method | Task A ($F_1$) | Task B (Recall) | Total |
|---|---|---|---|---|
| **ML4Hope**[4] | Random Forest | 0.62 | **0.39** | **1.01** |
| DreamTeam[5] | Random Forest | 0.39 | 0.28 | 0.68 |
| Bionedsnn[6] | Random Forest | **0.67** | - | 0.67 |
| team-23.5i[7] | MLP (?) | 0.40 | 0.00 | 0.40 |

The successful development of these algorithms has demonstrated the feasibility of making it possible to execute ML apps in a secure, decentralized environment without significant effort using FL techniques. Although this proof of concept has only one federated node, the same workflow is assumed to also be executable with multiple nodes [21].

### Workflow Feasibility

Although the model output did not show very high evaluation scores (see Table 1), they were able to generate output via the provided workflow.

This demonstrates that the workflow functioned as intended, enabling users to develop algorithms based on synthetic data and returned model statistics. This achievement underscores the feasibility of the proposed ecosystem for algorithm development and suggests that, given more time, participants could refine their models to achieve even better results.

[4] https://github.com/AleMer97/Makeathon-AMIGO
[5] https://github.com/LaurenzSommerlad/TUM.ai-Makeathon2024-Amigo-Challenge
[6] https://github.com/lkndl/makeathon/
[7] https://github.com/Warggr/ai-makeathon-23-5i

However, our workflow also revealed areas for improvement, as participants encountered some challenges during the TUMaiM24:

1. **Manual Testing**: The process of manually starting the FC workflow[8] on the real patient data by the challenge staff introduced delays. Automating this step will streamline the workflow and enable faster iterations in future events.
2. **System Compatibility**: Docker environment compatibility posed significant issues, particularly with containers created on Apple's ARM-based silicon architecture failing to run on the hospital's Linux servers. As this issue was found during the TUMaiM24, there was not enough time to search for the proper solution. Nonetheless, the buildx functionality of Docker, which has the capacity to build for both amd64 and arm64 concurrently, would have resolved the issue. Providing better documentation and support for different operating systems will mitigate these issues.
3. **Configuration Issues**: The time limit for graph queries was too strict, causing a timeout error if the query was more complex and took longer than a minute. This constraint restricted the participants to use simpler queries, hence impacting the retrieval of patient information from the graph for training. Addressing this limitation will be crucial to allow for more sophisticated algorithm development in the future.

In order to ensure the reproducibility of the entire setup and workflow, a fully containerized foundation[9] is provided. This process encompasses the development of applications. Subsequently, a validation script can be initiated with the results achieved in the application to get the model metrics for further model fine-tuning, as documented in the repository[10]. The simulated cKG access requirement is also accessible in the README of the same repository. The developed models and their application setup, tested with the workflow, can also be found in GitHub Repositories[11],[12],[13],[14].

## Comparison with Secure Processing Environment (SPE)

In order to meet the high regulatory requirements of the GDPR and those of the planned European Health Data Space [46], many private and public providers are reliant on a secure processing environment (SPE) [47–49]. This term refers to a local data enclave of a closed nature, in which sensitive health data remains within the data controller's firewall [49,50]. Analysts are only granted access to the data via highly secure virtual desktops. It is evident that all internet access is blocked, and all activities are subject to stringent monitoring [51]. The results are only released after rigorous testing, which may involve the use of an "airlock" to ensure compliance with the five security dimensions [52,53]. Given that raw data remains within the system, residual risks primarily stem from misuse by insiders or incorrectly configured export mechanisms [54].

[8] https://featurecloud.ai/assets/developer_documentation/workflow.html
[9] https://github.com/careforrare/PersonalizedMedicine
[10] https://github.com/careforrare/PersonalizedMedicine/tree/main/eval_metrics
[11] https://github.com/AleMer97/Makeathon-AMIGO
[12] https://github.com/LaurenzSommerlad/TUM.ai-Makeathon2024-Amigo-Challenge
[13] https://github.com/lkndl/makeathon/
[14] https://github.com/Warggr/ai-makeathon-23-5i

However, the approach presented here is more in line with FL, which pursues the opposite concept. Instead of exporting data, models leave the central unit as code packages, train locally, and exchange only encrypted model updates [55]. This eliminates the airlock waiting time, which significantly speeds up iterative development cycles. While SPEs scale by expanding local computing power, FL grows horizontally by integrating additional clients [55]. However, the increased agility leads to new risks, such as model inversion or gradient reconstruction [55]. These are mitigated by techniques such as secure aggregation, differential privacy, or trusted execution hardware [55,56].

In summary, it can be stated that although the principle presented here is associated with SPE, divergent approaches are being pursued, and comparability does not exist.

# Conclusion and Discussion

This study presents a proof-of-concept implementation of a privacy-preserving AI training pipeline in a single-client FL configuration, using a cKG for model prototyping and the FC infrastructure for secure deployment. This framework does not require developers to access sensitive patient data. Using cKGs to develop model prototypes and deploying models via the FeatureCloud infrastructure allowed external developers to design, refine, and evaluate algorithms in a controlled environment. The successful implementation of this approach during the TUM.ai Makeathon 2024 confirms its technical and operational feasibility.

This work highlights the value of simulated cKGs as semantically rich test environments that accurately reflect hospital data structures. These cKGs form a secure intermediate layer between conceptual model design and clinical deployment, enabling experimentation in full compliance with data protection regulations. The implemented models were purposefully kept simple. This was due to time constraints. Their seamless transition into a secure clinical environment underscores the potential of single-client FL. This is a preparatory phase for large-scale, multi-institutional consortia.

This study’s findings align with broader evidence in the literature that federated learning offers a promising avenue for medical AI development under stringent privacy regulations [20,23–25]. However, unlike most existing studies, which emphasize multi-client federated setups [26], this work highlights the utility of single-client FL deployments. These single-client setups provide a controlled environment for building and evaluating models without compromising data privacy. They represent a scalable intermediate stage, allowing researchers to build, test, and validate algorithms before expanding to full multi-institutional collaborations. The successful execution in a single-client setting demonstrates that the same workflow can be readily extended to multi-client environments.

Several technical limitations were identified during implementation, including container compatibility issues, strict query time constraints, and the necessity of manually executing evaluation scripts. These challenges demonstrate the need for further automation, standardized deployment templates, and improved infrastructure documentation to support more advanced model architectures and maintain development velocity. Overcoming these barriers is crucial for scaling FL-based pipelines beyond proof-of-concept implementations.

# Outlook

It is imperative that future research prioritize the optimization of existing processes, with a particular emphasis on the automation of test procedures. Additionally, there is a need to enhance interoperability between heterogeneous hospital environments and to expand the FL architecture to ensure increased scalability and efficiency. The expansion to encompass multiple clinics presents a valuable opportunity to utilize more extensive and diversified data sets, thereby facilitating the processing of complex issues and the analysis of diverse data types. This approach holds significant potential to not only substantially advance research in the field of rare diseases but also to enhance patient-centered medical care.

The overarching objective is to establish an infrastructure that facilitates secure access to sensitive medical data for ML development and industrial research without compromising data protection and ethical standards. The promotion of collaboration between academic institutions, hospitals, and industry is imperative to fully realize the potential of FL and achieve sustainable progress in medical research and care. The further development of FL systems is therefore a pivotal challenge that necessitates targeted investment and interdisciplinary collaboration to enable long-term innovations in personalized medicine.

In summary, it can be concluded that accelerated and optimized data access has the potential to enhance the efficiency of internal processes and to improve patient care. Targeted analysis of minimal data sets, such as individual blood values, facilitates the formation of informed decisions at an early stage. Concurrently, this development creates the opportunity to justify investments in more extensive and costly data generation processes, which may enable the development of more precise, personalized medicine.

# Acknowledgments

We would like to extend our deepest gratitude to Dr. von Hauner Children's Hospital for providing the multi-omics data access and for their commitment to pediatric research. Our sincere thanks go to the Care-for-Rare Foundation, LMU Klinikum, Capgemini, MI4People, and Neo4j for their support throughout this project. Special appreciation goes to TUM.ai for organizing the Makeathon and fostering an environment that encourages innovation and learning. We also thank all the participating students and professionals for their dedication and contributions to the TUM.ai Makeathon 2024 for advancing research on rare pediatric diseases. This work would not have been possible without the trust and consent of the parents and children, whose willingness to participate in research is helping to advance future medical breakthroughs. Funded by the European Union under contract no. 101136305. The Hungarian partner is funded by the Hungarian National Research, Development, and Innovation Fund. Views and opinions expressed are however, those of the author(s) only and do not necessarily reflect those of the European Union or the Hungarian National Research, Development and Innovation Fund. Neither the European Union nor the Hungarian National Research, Development and Innovation Fund can be held responsible for them.